\documentclass[aps,prc,showpacs,showkeys,notitlepage]{revtex4-1}
\usepackage[dvips]{graphicx}

\begin{document}

\title{{\footnotesize{Preprint of an article published in International Journal of Modern Physics E,  Vol. 25, No. 2 (2016) 1650008, \copyright World Scientific Publishing Company,  DOI: 10.1142/S0218301316500087}}\\ \bigskip Quantum diffraction grating.\\ A possible new description of nuclear elastic scattering.}

\author{H Wojciechowski\footnote{Emeritus, e-mail: Henryk.Wojciechowski@ifj.edu.pl} } 

\affiliation{The Henryk Niewodnicza\'nski Institute of Nuclear Physics, Polish Academy of Sciences, Radzikowskiego 152, 31-342 Krak\'ow, Poland }

\date{\today}

\begin{abstract}
  The problem of discontinuous functions and their representations in the form of Legendre polynomial series  in quantum nuclear scattering theory is  presented briefly. The problem is quite old yet not adequately explained in numerous Quantum Theory textbooks and sometimes not correctly understood by physicists. Introduction of the generalized functions into the quantum scattering theory clarifies the problem and allows to propose new interpretations of nuclear elastic scattering phenomenon. The derived new forms of the full elastic scattering amplitudes and possibility of splitting them suggest existence of dynamical quantum diffraction grating around the nuclei. Particularly important fact is that this grating existing in  the space around the nucleus makes considerable contribution to the experimental elastic differential cross section.  All these might be quite important in  analyses of nuclear elastic scattering data and so require to be stated in a more detailed and clear way.

\end{abstract}

\maketitle

\section{Introduction}

The nuclear scattering experiments (\textit{ie} the collisions of nuclear particles with the atomic nuclei)  are the fundamental experiments in nuclear physics. Among various types of scattering, the elastic scattering seems to be the simplest one.  In this experiment we have stream of free incident particles long before the collision with the scattering centre and again we have stream of free, the same as incident, scattered particles long after the collision.  After the collision  at large distance from the  scattering centre we count the number of particles scattered into certain solid angle $d\Omega$ at the angles ($\theta,\phi $) respectively to the direction of the incident beam. Assuming the axial symmetry of the process we can neglect the polar angle $\phi$  and so the whole process becomes a function  of azimuthal angle $\theta$ only. Counting the number of particles scattered at this angle we  construct the quantity named differential elastic cross section $\sigma_{el}(\theta)$  which we next compare with the predictions of various theoretical models.  If the incident particles are  spinless the differential elastic scattering cross section is the only quantity measured during elastic scattering experiments.  If the incident particles have spins, additionally to the former we can measure the polarization of the scattered particles \textit{ie}  the change in orientation of spins during the collision.  We do not  detect any signals from the scattering centre during the elastic collision so we  cannot to say  what is going on in the centre during the process of collision. The only information we  have is that the nucleus remains after the collision in its ground state. Some informations about the mechanism of the process we can get from the theoretical models describing the collision.  We  then use various theoretical  nuclear collision models and compare their predictions with the measured $\sigma_{el}(\theta)$.  The review of various possible theoretical treatments of nuclear scattering the reader can find in {\it The Quantum Theory of Nonrelativistic Collisions} by Taylor \cite{taylor1}.   

 Because of particle-wave dualism in quantum mechanics we can treat the stream of particles as the stream of wave packets  each of them having wave length $\lambda=h/p$ where $h$ is the Planck's constant and $p$ is the linear momentum of the packet  incident on scattering centre.  Due to this feature measuring the angular distributions of elastic scattered wave packets we observe the diffraction like structures similar to that observed in scattering of light.

 The simplest,  and the  historically first theoretical formalism for analysis of the scattering data is the time independent (stationary states) partial wave formalism.  In this picture  the incident stream of particles (wave packets) before the collision is represented by the plane wave, and after the collision the stream of  particles is represented again by the plane wave the same as before the collision plus the scattered spherical wave outgoing  from the scattering centre.  This is what we can read in every quantum scattering theory textbook. 

The simple phenomenological Optical Model (OM)  of nuclear elastic scattering based on the time independent partial wave formalism, so far, is the only model being able to reproduce easily and fairly well the measured elastic scattering differential cross sections for a wide range of scattered particles their energies and scattering centres (nuclei). The optical model has been developed many years ago \cite{hodgson} and is still successfully and commonly used for analyses of nuclear elastic scattering data.  Using this model and assuming the incoming particles are scattered on central potentials, Coulomb  $V_{Coul}(r)$ potential plus the nuclear complex potential $V_{nucl}(r) =  V(r) + iW(r)$, we can evaluate the function called the  elastic scattering   amplitude $f_{el}(\theta)$ which determines the probability  that particle is being scattered into an azimuthal angle $\theta$ on the reaction plane.  The predicted differential elastic cross section is just $\sigma^{th}_{el}(\theta) = |f_{el}(\theta)|^2$ which next is compared to measured one.

Almost identical with the partial wave picture of particle elastic scattering is the wave optics description of the scattering of light (\textit{ie} the stream of light quanta) by a small object \cite{hulst}.  Here we have two scenarios:

1.  If the plane wave front of light hits the scattering object which is completely absorbing (black)  after the collision a part of the plane wave front, in form and size of the object, is missing (absorbed by the object) and  the remaining incomplete wave front  forms a certain angular distribution of intensity called diffraction.

2.  If the scattering object is only partially absorbing,  additionally   to the diffraction patterns produced by incomplete plane wave front  we have light partly refracted (and reflected) by the object from that part of incident plane wave front which hits the object.  

In this pure optical description of scattering we have clearly defined the origin of the diffraction patterns, \textit{ie}, they are formed by the incomplete plane wave front which remains after the collision and which depend only on the size of of the object, the wavelength of the incident light, but does not depend on the opacity of it.  In this description after the collision we have incomplete plane wave front   producing diffraction, and additionally for opaque objects refracted wave outgoing from the scattering centre.  We see then, that there is ostensibly a small difference between the quantum mechanical partial wave picture of elastic scattering of nuclear particles by the nucleus and the  picture of scattering of light by a small object in the description what we have after the collision.

Below I would like to show  that in fact these two cases of scattering are almost identical, and then how adopting the above scattering of light picture to that of the elastic scattering of nuclear particles, allow us to separate  distinctly the nuclear   diffraction and nuclear refraction in the nuclear elastic scattering on the elastic scattering amplitude level. Such a separation might be very important and helpful in all kinds of analyses of nuclear elastic scattering data.  In the present paper I will discuss only the simplest case of elastic scattering neglecting the spins  of incident particles.

\section{Some elemental theory}

It has been pointed out by Maj and Mr\'owczy\'nski \cite{maj} that in this simple time independent partial wave formalism of nuclear elastic scattering we use some mathematical relations which, from the mathematical point of view, seem to be inconsistent but which give correct final results . 

First of all we must overcome these, mentioned above, "inconsistencies" of some mathematical relations present in this formalism.  This can easily be done by invoking here  long time ago introduced by Schwartz \cite{schwartz} the generalized functions (distributions).

Let us have a look at just few examples of relations used in partial wave picture of nuclear scattering theory which seem to be "mathematically inconsistent".  The first example is the plane wave (or rather its asymptotic form at large distance) decomposition into spherical waves  which is

\begin{equation}
  e^{ikrcos\theta} = \frac{1}{kr}\sum\limits_{l=0}^\infty i^l(2l+1)P_l(cos\theta)sin(kr-\frac{\pi l}{2}).
\end{equation}

The second is the commonly used partial wave "decomposition" into Legendre polynomial series  of the analytical form of Coulomb scattering amplitude (\textit{ie} the scattering amplitude for Coulomb potential)

\begin{eqnarray}
\frac{n}{2k\sin^2{\frac{1}{2}}\theta}\exp({-in\ln(\sin^2\frac{1}{2}\theta)+i\pi+2i\eta_0}) = & \nonumber \\ \frac{1}{2ik}\sum\limits_{l=0}^\infty (2l+1)(S_l^{Coul}-1)P_l(cos\theta).
\end{eqnarray}

There is also a third decomposition of the well known Dirac's $\delta $-function into spherical waves, which is \cite{mott}

\begin{equation}
\delta (1-cos(\theta) ) = {\frac{1}{2}} \sum\limits_{l=0}^\infty (2l+1)P_l(cos\theta),
\end{equation}

where in all above relations: $\theta $ is the azimuthal angle in spherical coordinate system, $l$ is the angular momentum quantum number, $P_l(cos\theta )$ are the Legendre polynomials of $l$ order, and $k$ is the wave number, $n$ is the Sommerfeld parameter, the $S_l^{Coul}=e^{2i\eta _l}$ in  (2) are the Coulomb scattering matrix elements and $\eta_l = arg\Gamma(l+1+in)$.

Function at left hand side in  (1), the plane wave, represents a stream of incident particles, that is,  represents the case of an ideal parallel beam of particles moving all along the "z" axis (or at zero degree in spherical coordinate system). It represents then the "point" angular distribution of the incident beam.

Function at left hand side in (2) is the analytical Coulomb scattering amplitude for the scattering of two point charged particles and  is well known to have a singularity at zero degree.

The Dirac's $\delta $-function in (3) is undefined (or sometimes defined as infinite) at zero degree and is zero for all other angles.  Again we have a function which has the point distribution.

All series at right-hand sides in (1), (2) and (3) are diverging at the point of $cos\theta =1$, and the equality signs in these relations seem to be then meaningless.    Although these equations seem to be incorrect, the final results obtained by means of them are correct. These expressions, especially the first one, are present in all textbooks of Quantum Mechanics.  This problem has been widely discussed in papers of  Maj and  Mrowczynski {\cite{maj}} and of Marquez {\cite{marquez}}, nevertheless, both papers do not give correct explanation of it.

In order to treat such discontinuous functions more like smooth functions, the concept of generalized functions (distributions) has been introduced by   Schwartz in 1950 {\cite{schwartz}}. The generalized functions allow us to treat such problems easily in mathematical way.  Albert Messiah in his {\it Quantum Mechanics}, Vol. 1, Appendix A {\cite{messiah}}, introduced an elements of distributions theory  but mainly in its aspect to the $\delta $-function.

If we then treat the relations (1), (2) and (3) as normal  functional relations the equality signs in these equations are meaningless. If we regard them as generalized functions the equality signs are perfectly correct since the right hand sides series, as distributions, converge weakly {\cite{web}} to the left hand sides functions in the whole angular range including zero degree. It is then clear that presenting the above relation it should be clearly stated that we are dealing with generalized functions.

Let us examine the second of the above relations, that is the Legendre polynomial "expansion" of the analytical Coulomb scattering amplitude.  This case has been widely discussed by Taylor in his paper published in 1974 {\cite{taylor}}. 
He proved there, that the analytical form of Coulomb scattering amplitude cannot be expanded into Legendre series, since the obtained series is divergent.  It can, however, be replaced by  Legendre polynomial series, which, as distribution, weakly converges to it. Moreover, he shows that any of two such series, which coefficients of expansion differ  by an arbitrary  independent of $l$ constant, also weakly converge to it. Invoking generalized functions we got then infinite number of series converging weakly to the analytical Coulomb amplitude. We see then that the unique expansion of Coulomb scattering amplitude into Legendre polynomial series does not exist. 

Schiff in his {\it Quantum Mechanics},  starting from the third edition  {\cite{schiff}}, presents another version of relation (2). He writes 

\begin{equation}
%  \begin{eqnarray}
\frac{n}{2k\sin^2{\frac{1}{2}}\theta}\exp({-in\ln(\sin^2\frac{1}{2}\theta)+i\pi+2i\eta_0}) =  \frac{1}{2ik}\sum\limits_{l=0}^\infty (2l+1)S_l^{Coul}P_l(cos\theta).
%\end{eqnarray}
\end{equation}

He writes that this is a Legendre polynomial expansion of the Coulomb scattering amplitude, which, according to Taylor {\cite{taylor}} as a unique expansion does not exist. The right hand side series of above can only be regarded  just as one of the series presented by Taylor where the arbitrary constant  equals zero.

In some quantum mechanics textbooks {\cite{landau}} the above relation is sometimes treated as a special case of relation (2) for $\theta \neq 0$.  In such a case the above relation should vanish when $S_l^{Coul} =1$.

Schiff did not mention explicitly that he deals with generalized functions but he writes next that for  $S_l^{Coul}=1$ this becomes 

\begin{equation}
%\begin{eqnarray}
 \frac{1}{2ik}\sum\limits_{l=0}^\infty (2l+1)S_l^{Coul}P_l(cos\theta)  \rightarrow  \frac{1}{2ik}\sum\limits_{l=0}^\infty (2l+1)P_l(cos\theta).
%\end{eqnarray}
\end{equation}

which behave like $\delta $-function.

 It in fact means, that he treats his relation in a sense of generalized functions. (One should notice that series in relations (2) and (4) differ only at $\theta = 0$, \textit{ie}, at the point where the experimental elastic differential cross section is meaningless since one cannot distinguish there the particle that is scattered from  the particle that is not scattered.)

This fact is very important, since relations (2) and (4) are fully equivalent (even if they differ by the $\delta $-distribution) independent representations  of the analytical Coulomb scattering amplitude  in the sense of distributions. Physics reduced those infinite number of series mentioned by Taylor to only two where the arbitrary constant can be $0$ or $-1$. 

Fact that in partial wave picture of quantum scattering theory we have two series fully representing the analytical Coulomb scattering amplitude might be a kind of surprise.   

 The full scattering amplitude for elastic scattering of charged particles by Coulomb and nuclear forces which is usually presented  as

\begin{eqnarray}
f_{el}(\theta )=\frac{n}{2k\sin^2{\frac{1}{2}}\theta}\exp({-in\ln(\sin^2\frac{1}{2}\theta)+i\pi+2i\eta_0}) + \nonumber \\ \frac{1}{2ik}\sum\limits_{l=0}^\infty (2l+1)S_l^{Coul}(S_l-1)P_l(cos\theta).
\end{eqnarray}

now can be reduced to its mathematical simplest form  of only one simple series \cite{hw1} 

\begin{equation}
%\begin{eqnarray}
\fbox{$f_{el}(\theta) =  \frac{1}{2ik}\sum\limits_0^\infty (2l+1)S_l^{Coul} S_l P_l(\cos\theta)$}
%\end{eqnarray}
\end{equation}

which for the case of neutral particles ($S_l^{Coul}=1$)  reduces to

\begin{equation}
\fbox{$f_{el}(\theta)=\frac{1}{2ik}\sum\limits_0^\infty (2l+1)S_l P_l(\cos\theta)$}
\end{equation}

The above series are divergent at zero degree and very slowly convergent for all other angles.  They are then useless directly for evaluations of differential cross sections but they are perfect for interpretation of the elastic scattering phenomenon. In fact relations (6) and (7) are two forms of the same relation.  We can however accelerate they convergence using series comparison method \cite{age} and doing that we instantly receive well known relation (6) commonly used for evaluations of differential cross sections.  Series (7) and (8) while divergent at forward angle can be, as it will be shown later, integrated over whole angular range in certain situation.

\section{The sum-of-differences (SOD) formula}

The presented forms (7) and (8) of the elastic scattering amplitude seem to be not acceptable  since they cannot be directly used for evaluations of differential cross sections and cannot be directly integrated over the whole angular range to get the total integrated elastic amplitude.  This is not exactly true.  Indeed, the above series are  divergent  at zero degree  but the differential elastic cross sections calculated by means of them can be integrated under certain conditions.  Let us assume  that we can switch-off the nuclear forces for the moment ($S_l =1$) leaving only Coulomb forces, then the elastic differential scattering cross section should be $ \sigma_{el}(\theta) = |f_C(\theta)|^2$ calculated using the right hand side series of relation (4).  If we now switch-on the nuclear forces then the elastic differential scattering cross section will be $ \sigma_{el}(\theta) = |f_{el}(\theta)|^2$ and will be calculated using series from   relation (7).  The missing flux is then the integral over the whole angular range of the difference of the above cross sections.  This integral can easily be evaluated and the result is \cite{hw2}

\begin{equation}
 \fbox{$ 2{\pi}\int\limits_{0}^{\pi}\{|f_C(\theta)|^2 - |f_{el}(\theta)|^2\} \sin{\theta} \; d\theta = \pi/k^{2}\sum\limits_{l=0}^{\infty}(2l+1)(1-\mid{S_l}\mid^2) = \sigma_r $}
\end{equation}
\medskip

  Moreover, the same can be done for the case of neutral particles where $S_l^{Coul}=1$ and again the result is $\sigma _r$.  It is then clear that the amplitudes (7) and (8) are perfectly flux conserving.   These amplitudes cannot be integrated separately but the differences of the elastic differential cross sections calculated using them are perfectly integrable.

The variation of the above relation, named sum-of-differences (SOD), has been proposed early \cite{davis} for evaluation of the total reaction cross sections from the experimental elastic differential cross sections for strongly absorbed  particles in the presence of strong Coulomb field (the "Fresnel type" of elastic angular distributions) is

\begin{equation}
 SOD(\theta_{st}) = 2{\pi}\int\limits_{\theta_{st}}^{\pi}\{|f_C(\theta)|^2 - |f_{el}(\theta)|^2\} \sin{\theta} \; d\theta = \sigma_r
\end{equation}

where $\theta_{st}$ is the starting angle for integration (summation) where $\sigma _{el}/\sigma _{Ruth}$ oscillate around one.

 The comparison of the extracted $\sigma _r^{SOD}$ with that predicted by optical model (OM) was satisfactory.  
It was found that modified form of the SOD formula, the MSOD($\theta_{st}$) \cite{hw2,marty} as a function of $\theta_{st}$ oscillates around $\sigma _r$ at forward angles for "Fresnel" type angular distributions and it can be used for the model independent method of evaluation of  $\sigma _r$  from experimental data.

\section{Separation of diffraction and refraction in angular momentum space}

The above (7) and (8) forms  of full elastic scattering amplitudes might lead to the very simple  interpretations of the elastic scattering phenomenon. 

Initially let us assume that the scattering object and incident particle have well defined sharp surfaces.  In this situation  the nuclear scattering matrix elements ${\mid S_l^{nucl}\mid\leq 1}$ for angular momenta below certain grazing  momentum $l_{gr}$ and ${\mid S_l^{nucl}\mid =1}$ above $l_{gr}$.  (We have to notice that above $l_{gr}$ we have only $Re S_l^{nucl}$ because $Im S_l^{nucl}$  are equal zero).

In this case  the above amplitudes can easily be split into two parts, low angular momenta $l\leq l_{gr}$ part  and high angular momenta $l>l_{gr}$  part:
\begin{equation}
%\begin{eqnarray}
f_{el}(\theta) =  \frac{1}{2ik}\sum\limits_0^{l_{gr}} (2l+1)S_l^{Coul} S_l P_l(\cos\theta) + 
\frac{1}{2ik}\sum\limits_{l_{gr}+1}^\infty (2l+1)S_l^{Coul}P_l(\cos\theta)
%\end{eqnarray}
\end{equation}

and for neutral particles

\begin{equation}
%\begin{eqnarray}
f_{el}(\theta) =  \frac{1}{2ik}\sum\limits_0^{l_{gr}} (2l+1)S_l P_l(\cos\theta) + 
\frac{1}{2ik}\sum\limits_{l_{gr}+1}^\infty (2l+1)P_l(\cos\theta)
%\end{eqnarray}
\end{equation}

If we now invoke the cited earlier definition of diffraction \cite{hulst} we see that the second series at right hand sides of relations (11) and (12) describe just nothing but pure diffractions on black or opaque nucleus of "radius" $l_{gr}$  in angular momentum space  which are independent of nuclear scattering matrix elements!  The first series there, dependent  on nuclear scattering matrix elements, describe nuclear refraction on opaque nucleus of the same "radius" $l_{gr}$. 

The schematic scenario of the above situation  (for neutral particle case) is presented in Fig. 1

\begin{figure}[htb]
\begin{center}
\includegraphics[width=7cm]{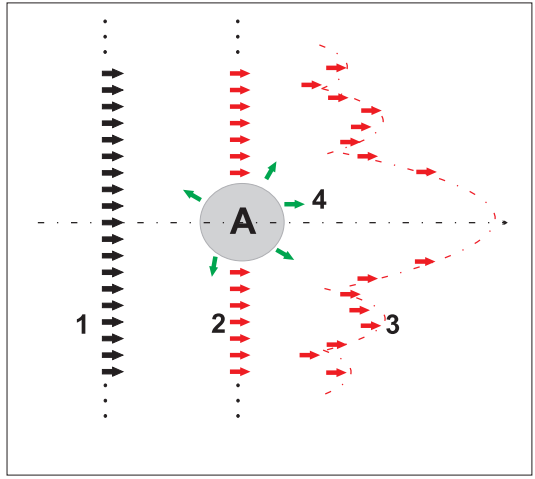} 
\caption{A - scattering centre (black or opaque nucleus), 1 - incident plain wave front (beam of incident particles), 2 - incomplete plane wave front (incident particles which passed by the nucleus), 3 - observed diffraction patterns formed by the incomplete plane wave front, 4 - refracted and reflected particles which hit the opaque nucleus. }
\end{center}
\end{figure}

If we assume  that the nucleus is completely black (all $S_l=0$ below $l_{gr}$) we have only second series in relations (11) and (12) and we have nothing but the old sharp cut-off Blair diffraction model amplitudes for scattering by black nucleus \cite{blair}  not only for charged but also for neutral particles. 

The above decompositions of amplitudes (11) and (12) is equivalent to decompositions of the nuclear scattering matrix elements $S_l$ into refractive and diffractive parts
{
\begin{eqnarray}
Re{\mathcal{S}}_l^{ref} = ReS_l & - {\mathcal{F}}_l^{diff} \\ 
Im{\mathcal{S}}_l^{ref} = ImS_l & \nonumber
\end{eqnarray}}

where in the sharp cut-off case the $Re{\mathcal{S}}_l^{ref}\equiv ReS_l $ below $l_{gr}$ and equal zero above it, the ${\mathcal{F}}_l^{diff}$ are zero below $l_{gr}$ and are equal one above it while the imaginary parts remain unchanged $ImS_l \equiv Im{\mathcal{S}}_l^{ref}$ below $l_{gr}$ and vanish above it.
\medskip

The separation of the elastic scattering amplitude (11) can then be written generally  as
\begin{equation}
\fbox{$f_{el}(\theta) =  \frac{1}{2ik}\sum\limits_0^\infty (2l+1)S_l^{Coul}\mathcal{S}_l P_l(\cos\theta) + 
\frac{1}{2ik}\sum\limits_{0}^\infty (2l+1)S_l^{Coul} \mathcal{F}_l P_l(\cos\theta)$}
\end{equation}
\bigskip
 
Let us look now at the second series in relation (12). Using relation (3) we can write that
\begin{equation}
%\begin{eqnarray}
\frac{1}{2ik}\sum\limits_{l_{gr}+1}^\infty (2l+1)P_l(\cos\theta)= -\frac{1}{2ik}\sum\limits_{0}^{l_{gr}} (2l+1)P_l(\cos\theta)
%\end{eqnarray}
\end{equation}

The above relation is nothing but the Babinet principle which states that the diffraction (patterns) from a black body of "radius" $l_{gr}$ is identical to that from a hole in black screen with the same "radius" in angular momentum space {\cite{barone}}.  

Replacing now the last series in relation (12) by the right hand side series of relation (15) we receive well known  scattering amplitude for neutral particles of the form

\begin{equation}
f_{el}(\theta) =  \frac{1}{2ik}\sum\limits_0^{l_{gr}} (2l+1)(S_l -1) P_l(\cos\theta) 
\end{equation}

The optical Babinet principle holds for the incident plane wave, nevertheless we can write relation equivalent to (15) for charged particles where we have Coulomb distorted plane wave, and in sharp cut-off case, we have

\begin{eqnarray}
\frac{n}{2k\sin^2{\frac{1}{2}}\theta}\exp({-in\ln(\sin^2\frac{1}{2}\theta)+i\pi+2i\eta_0}) - \frac{1}{2ik}\sum\limits_{0}^{l_{gr}} (2l+1)S_l^{Coul}P_l(cos\theta) \nonumber \\ = \frac{1}{2ik}\sum\limits_{l_{gr}+1}^\infty (2l+1)S_l^{Coul}P_l(cos\theta) .
\end{eqnarray}

In reality the nucleus has diffused edge and the separations of (11) and (12) must be done in a smooth cut-off case.  
This can be done by  using a discrete values for  ${\mathcal{F}}_l^{diff}$  of a smooth function  $F(l)$ gradually varying from zero to one around $l_{gr}$.  The $F(l)$ function (here, only for illustrating the method, it is taken from the semi-classical models) can be of the following  shape:

\begin{equation}
{F}(l) = \Bigg{\lbrack} 1+exp{\frac{l_{gr}-l}{\Delta}}\Bigg{\rbrack} ^{-1}
\end{equation}

The parameters of the above $F(l)$ function, \textit{ie},   the $l_{gr}$ and $\Delta$  can be obtained by fitting the tail of $ReS_l$ with $F(l)$.  The details of this one can find in previously cited paper {\cite{hw1}}.  The imaginary parts of $S_l$ are left untouched. 

Illustration   of such a  smooth cut-off separation of the $ReS_l$ into $Re{\mathcal{S}}_l^{ref}$  and ${\mathcal{F}}_l^{diff}$ is presented in Fig 2. 

\begin{figure}[htb]
\begin{center}
\includegraphics[width=7cm]{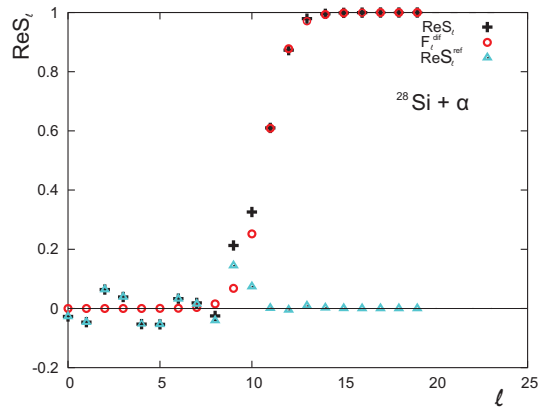} 
\caption{ The smooth cut-off decomposition of the $ReS_l$ into $Re{\mathcal{S}}_l^{ref}$  and ${\mathcal{F}}_l^{diff}$  for $\rm{^{28}Si+\alpha}$ elastic scattering  at 26.5~~MeV  incident energy.}
\end{center}
\end{figure}

The ${\mathcal{F}}_l^{diff}$ coefficients in (13) and (14) are real and can be regarded as dumping factors for Coulomb scattering amplitude inside the tail of nuclear interaction region.  Much more realistic way of calculating ${\mathcal{F}}_l^{diff}$ coefficients would be, for instance,  by evaluation of  ${F}(l)$ function from charge distribution of individual nuclei.

\section{Quantum diffraction grating}

The main question is, if the cited above classical wave optics definition of diffraction \cite{hulst} is suitable for the case of the nuclear wave packets  elastic  scattering, \textit{ie}, do really incomplete plane wave front passing by the nucleus is responsible for  creation of diffraction patterns?  Let us then examine in more detail relation (12).  

The partial wave formalism of elastic nuclear scattering used here has one very important aspect.  The angular momentum quantum numbers $l$ present in this formalism can have only integer values, \textit{ie}, $l$ can be $0, 1, 2, 3, ...$.  Relations (7) and (8) are the sum of discrete partial wave probabilities of scattering for each angular momentum quantum number $l$. The angular momentum of the incoming packet $L$ can be expressed by means of angular momentum quantum number by simple formula

\begin{eqnarray}
L = \sqrt{l(l+1)}\hbar \nonumber
\end{eqnarray}
which can be related to the impact parameter (the distance of a particle trajectory from the scattering centre) $b$ by the relation

\begin{eqnarray}
L = bk  \nonumber
\end{eqnarray}
where k is wave number.

Due to quantization $l$ can  have only an integer value and  then the $L$ and impact parameter $b$ must  have also discrete values $L_l$ and $b_l$.   Incident wave packets with impact parameters $b \leq b_{l_{gr}}$ enter the nuclear interaction region while those having $b > b_{l_{gr}}$ passing by the nucleus are directed into the discrete channels $b_l$ with $l>l_{gr}$. 

Quantization create in the space around the nucleus a specific set of discrete impact parameters circles (scattering is axially symmetric around the incident beam direction)  which are the only "channels" in space allowed to incident packets while passing by the nucleus.  

The channels up to $l_{gr}$ are shaded by nucleus (scattering centre) while channels above $l_{gr}$  are open.  Above $l_{gr}$  we have nothing but the circular quantum dynamical diffraction grating formed by quantization around the scattering centre where "slits" of grating are placed at the position $b_l$ for quantum numbers $l=  l_{gr}+1, l_{gr}+2 .....$. Fig. 3 illustrates this on a plane perpendicular to the beam axis. This clearly shows that wave packets passing by the nuclear region must go through quantum diffraction grating and really undergo diffraction creating diffraction patterns. Relation (12) is for scattering of neutral particles but the same situation we have for positively charged particles (relation (11)) since there the Coulomb repulsive forces outside the nucleus act only like a diverging lens distorting  the incomplete plane wave outside the nucleus. The width of the slits  depends on the energy spread of incident beam.  The minimal width for strictly monoenergetic particle is limited by the Heisenberg uncertainty principle. For attractive Coulomb force (electrons beam)  the "slits" are converted into spherical  shells around the nucleus where electrons can be captured forming (under suitable condition) stable atomic orbits.

\begin{figure}[htb]
\begin{center}
\includegraphics[width=7cm]{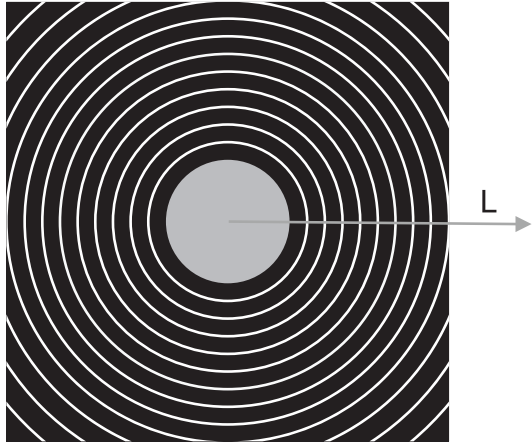} 
\caption{Schematic view of quantum diffraction grating. Central opaque area represent the nucleus (area of nuclear forces), the white circles around it represents the open channels (slits) of the quantum grating.}
\end{center}
\end{figure}

The effect measured as elastic nuclear scattering is then the sum of refracting and diffracting effects and theoretically is  expressed  by relation (14) . It is then clear that diffraction is present only in nuclear elastic scattering channel.

\section{Separation of the measured elastic cross sections into diffractive and refractive parts}

The example of separation of the elastic differential cross sections (Optical Model) for $\rm{^{28}Si+\alpha} $ at $\rm{E_{\alpha }=26.5}$MeV for the decomposition of $ReS_l$ as shown in Fig. 2 is presented in Fig. 4.

\begin{figure}[htb]
\begin{center}
\includegraphics[width=10cm]{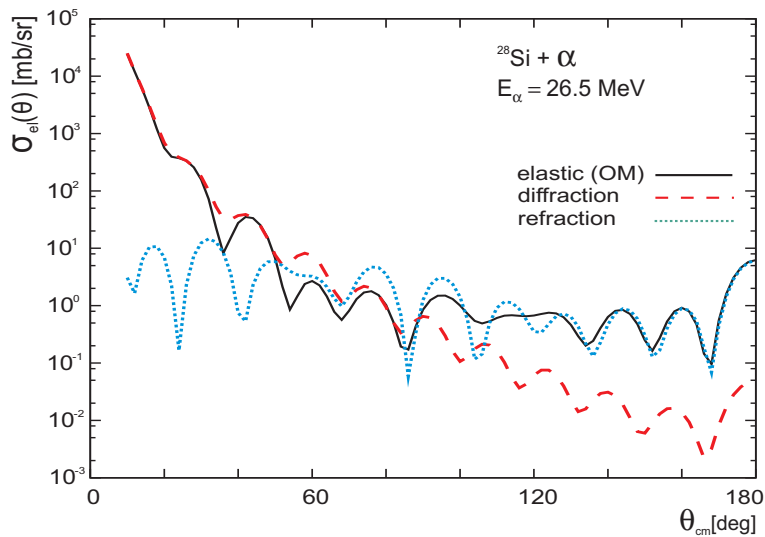} 
\caption{ The decomposition of the optical model (solid line) differential
cross sections into diffractive parts (dashed line), calculated using 
diffractive scattering amplitude, and refractive parts
(dotted line), calculated using refractive scattering amplitude, 
for $\rm{^{28}Si+\alpha }$ at 26.5 MeV elastic scattering.}
\end{center}
\end{figure}

The same separations for various nuclear scattering systems and scattering energies, published elsewhere and cited early \cite{hw1} are presented in Fig. 5. 

\begin{figure}[tbp]
\begin{center}
\includegraphics[width=10cm,height=20cm]{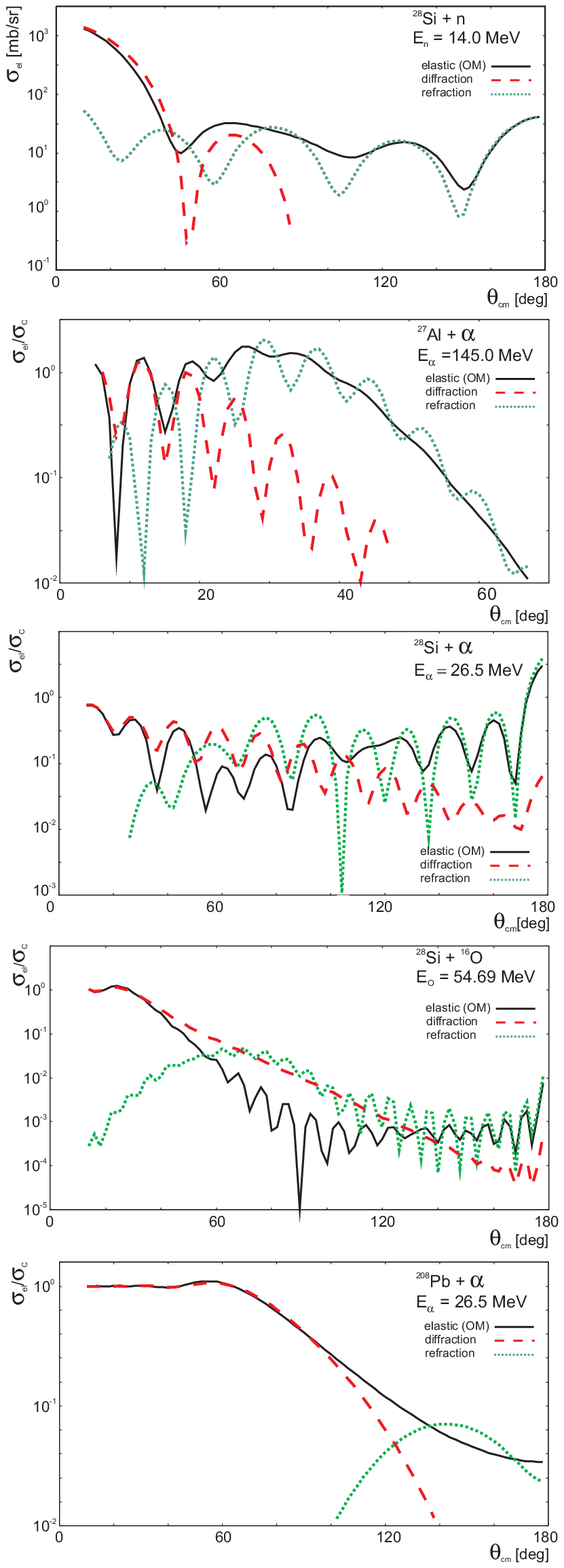} 
\caption{ The decomposition of the optical model (solid line) differential
cross sections into diffractive parts (dashed line)  and refractive parts
(dotted line) for various nuclear scattering systems.} 
\end{center}
\end{figure}

Looking at Figs 4 and 5 we see that at the backward scattering angles diffraction and refraction cross sections are well separated in most  presented low energy scattering data and that the refraction dominates there.

%\newpage
\section{The Optical Model analyses of the nuclear elastic scattering data and the goodness-to-fit test function}

The fact that the nuclear elastic scattering can be easily decomposed into diffractive and refractive parts which have  different origin  in angular momentum space and fact that these two components are dominant at  different angular regions in experimental differential cross sections (see Figs. 4 and 5), should  be  a crucial point in analyses of experimental elastic data.

The analyses of the nuclear elastic scattering differential cross sections are usually  performed using the nuclear optical model (OM), in which the nuclear scattering matrix elements are evaluated by solving the Schr\"odinger equation for the sum of Coulomb potential $V_C(r)$ and complex nuclear potential $V_n(r)= U(r)+iW(r)$, and then the differential elastic cross section is calculated by means of elastic scattering amplitude of form (6). The whole calculations are performed numerically on a computer, and the calculated elastic differential cross sections $\sigma _{OM}(\theta_i ) = |f_{el}(\theta_i )|^2$ are  compared to that measured experimentally using  the so called "goodness-to-fit" test function.

As a test function the "approximate"  $\chi ^2$ test function has been proposed by Hodgson {\cite{hodgson}}  and it has the form: 

\begin{equation}
\chi ^2 \approx  \Delta = \sum_{i=1}^n \left( \frac {\sigma _{th}(\theta _i) - \sigma _{exp}(\theta _i)}{\delta \sigma _{exp}(\theta _i)}\right) ^2 
\end{equation}

where: $\sigma _{th}(\theta _i)$ is the predicted cross section for angle $(\theta _i)$, $\sigma _{exp}(\theta _i)$ is the experimentally measured cross section for the same angle, and summation is over all angles $(\theta _i)$ of measured differential cross section. 

The above form of the "goodness-to-fit" test function has been widely accepted and is commonly in use.  

Computer adjusts the optical model starting parameters as long as the test function reaches its minimum. This procedure is called an automatic search routine and usually is based on the gradient of goodness-to-fit function in the OM parameters space. It seems that any smooth function which have well defined   minimum when all $ {\sigma _{th}(\theta _i) = \sigma _{exp}(\theta _i)}$
can be used here.  This assumption generally is not true!  This is true only when we analyse  the data of a single phenomenon.  Here, as has been shown above, we have data of two different phenomena - diffraction and refraction which are of completely different nature and origin.  Here we should use correct statistical $\chi ^2$  test function instead of its approximate version.

The  correct  $\chi ^2$  (derived from the statistical $\chi ^2$) test function for analysis of nuclear scattering data {\cite{hodgson}} is :

\begin{equation}
\chi ^2 = \sum_{i=1}^n \left( \frac {\sigma _{th}(\theta _i) - \sigma _{exp}(\theta _i)}{\delta \sigma _{exp}(\theta _i)}\right ) ^2\frac {\sigma _{exp}(\theta _i)}{\sigma _{th}(\theta _i)}
\end{equation}

Functions (19) and (20) have theoretical minima at the same point but in fact they are  completely different function.  To see the difference  between them, we can plot one component of each as a function of the $\sigma _{th}/\sigma _{exp}$ ratio.  Fig. 6 shows the difference between them.

\begin{figure}[htb]
\begin{center}
\includegraphics[width=7cm,height=5cm]{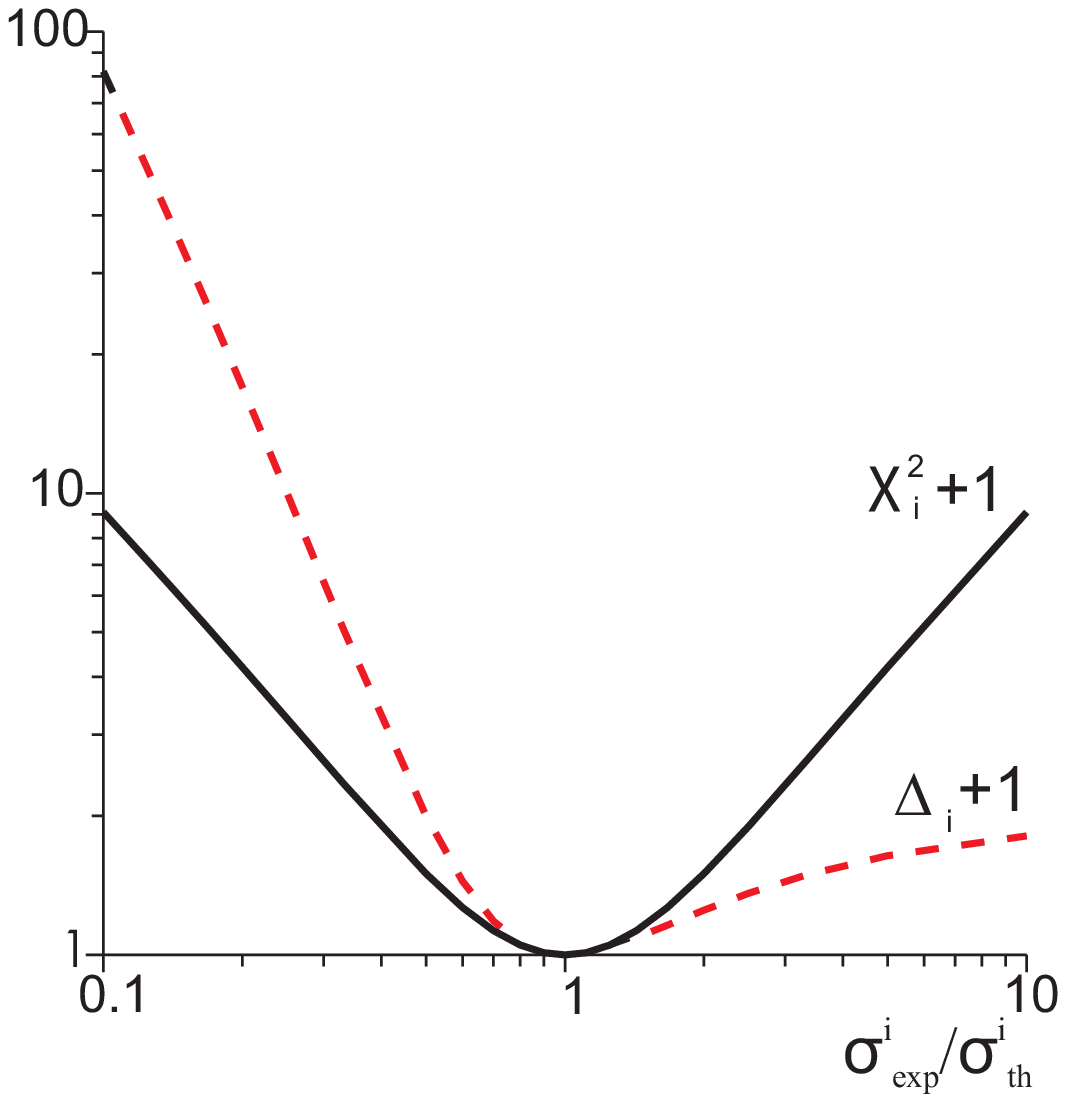} 
\caption{ The plot of $\chi _i^2$ and $\Delta _i$ single components of total $\chi^2$ and $\Delta$ as a function of $\sigma _{exp}/\sigma _{th}$ ratio.  }
\end{center}
\end{figure}

In optical model the  $\sigma _{exp}/\sigma _{th}$  ratio at backward angles of differential elastic cross sections is controlled by imaginary part of the nuclear optical potential $W(r)$. If we now look at the profiles of the total correct $\chi _i^2$ and its approximate version $\Delta _i$ minima along the $W(r)$ lines shown in  Fig. 7, we see the same as presented in Fig. 6 behaviour.  

\begin{figure}[htb]
\begin{center}
\includegraphics[width=7cm,height=5cm]{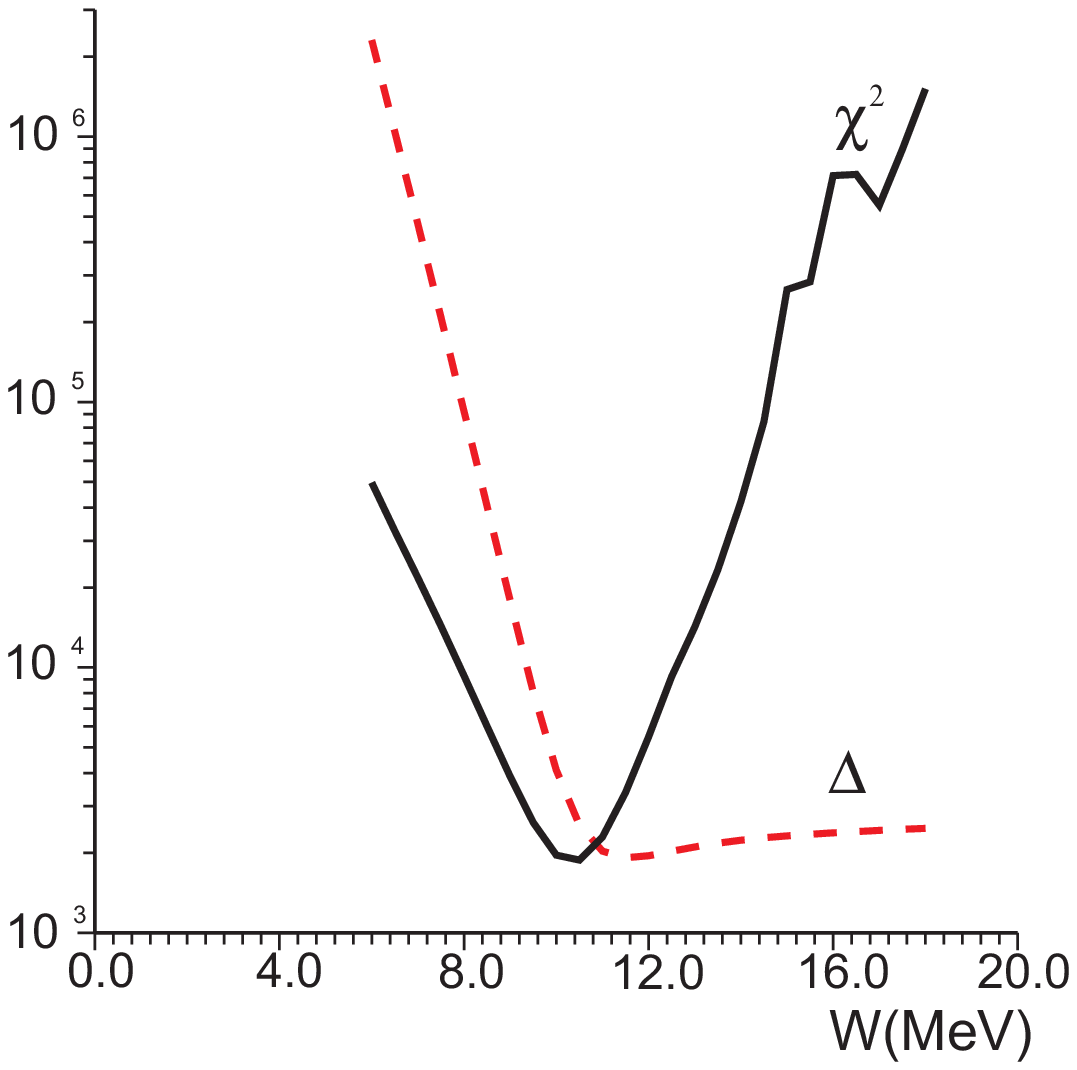} 
\caption{ The profiles of $\chi^2$ and $\Delta$ test functions along the $W(r)$ lines for minimum of optical model parameters for the case of $\rm{^{40}Ca + \alpha} $ at $\rm{E_\alpha = 26 MeV}$ using true $\chi ^2$ and $\Delta $ tests.}
\end{center}
\end{figure}

The imaginary part of the nuclear potential $W(r)$ is responsible for nuclear absorption and so influences only on refracted cross section which is dominant at backward angles.  Due to the flatness of the right branch of  $\Delta$ along $W(r)$ axis the search usually gives wrong fits at backward angles.

\begin{figure}[htb] 
\begin{center}
\includegraphics[width=10cm]{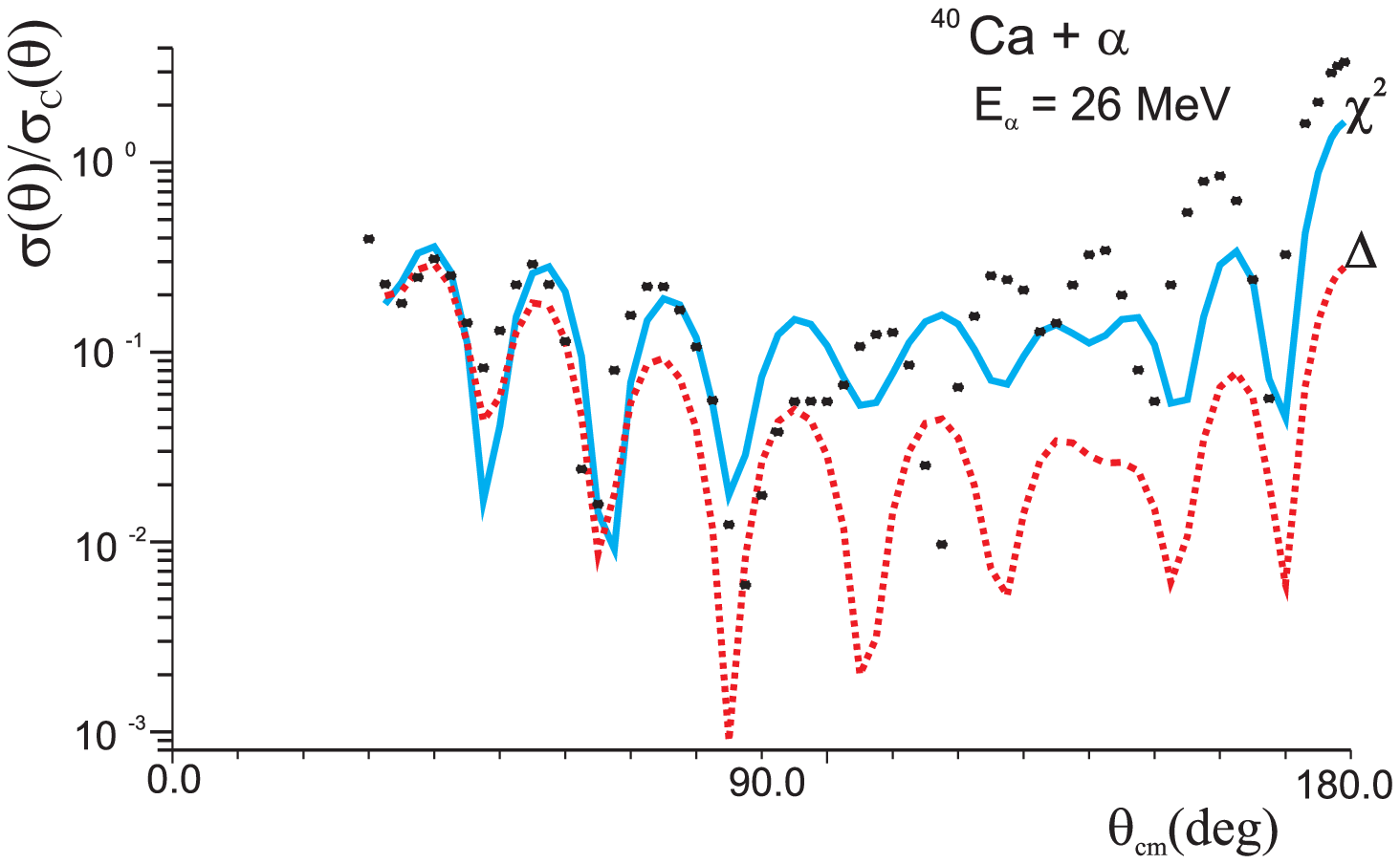} 
\caption{Results of the final fits to the experimental data of $\rm{^{40}Ca + \alpha} $ using true $\chi ^2$ and $\Delta $ tests.  }
\end{center}
\end{figure}

Fig. 8 illustrates  the quality of fits using true $\chi ^2$ and $\Delta $ (\textit{ie} the approximate $\chi ^2$) tests for just one example of elastic nuclear differential cross section.
Final optical model parameters determined by an automatic search routine and by scanning the test functions surfaces,  not only for this scattering system but also for many other, are also completely different \cite{hw3}.  Here again has to be pointed out that this is due to the fitting of two different phenomena.

\section{Summary}

Extending the commonly used in scattering theory mathematical formalism to the generalized functions leads to the much simpler and transparent for interpretation description of the nuclear elastic scattering phenomenon. As one can see above, treating only the well known analytical formula for Coulomb scattering amplitudes by means of generalized functions gives us a very simple formulae for the elastic scattering amplitude for charged and neutral particles.  These simple formulae of elastic scattering amplitudes can be very easy split in angular momentum space into two parts,  low angular momentum refraction part for $l \leq l_{gr}$ and high angular momentum diffraction part for $l>l_{gr}$.

This picture and presented above separation of diffractive and refractive effects for few measured elastic scattering data (experimental data are the same as used in previous publications \cite{hw1,hw3}) clearly show that the nuclear elastic differential cross sections are mainly of diffractive nature and that the nuclear effects are very small and visible only at backward angles. Presented above separations of the elastic scattering amplitudes reveal existence  of dynamical diffracting  grating outside the nucleus which  diffract the passing-by particles giving diffractive effects which is much larger than that given by the nucleus.   This is nothing unusual since the        surface of nuclear refracting region extending for $l$ from $0$ to $l_{gr}$ is much smaller than the surface of the outer quantum diffraction grating which extends for $l$ from $l_{gr}$ to infinity. This can easily be illustrated by integrating the measured elastic differential cross section over an angular range of measurement and the refracted differential cross section calculated using low angular momenta scattering amplitude we will see that the latter  is only a very small fraction  of the whole phenomenon. If we do this for the case of scattering of alpha  particles on $\rm{^{28}Si }$ nuclei at $\rm{E_{\alpha} = 26.5 MeV} $ presented in Fig 4 over an angular range from $10^0$ to $180^0$ we find that the ratio of refraction to total is only $4.46$ x  $10^{-3}$ . If we then want to get correct information about nuclear interaction  we must analyse the whole differential measured elastic cross section by nuclear interaction model (here the Optical Model) including the backward angle  part of it  where the refraction dominates.  In many cases of scattering of strong absorbed particles the cross sections at backward angles are so low that are unmeasurable and our measurements and analyses are limited only to forward angles where diffraction  highly dominates.  All such analyses are  rather of a very low value.  

Even if we have data covering the backward angles where the refraction  dominates    and if we use computer programme with automatic search routine for comparison of predicted cross section to that of measured it is extremely important to use appropriate goodness-to-fit test function.  In Fig 8 we see that two different test 
functions can produce different fits even if both functions have minima at the same point, \textit{ie},  at the point where measured cross section has the same value as predicted.
These difficulties arise only in elastic nuclear reaction channels and are due to the fact that we are dealing with two phenomena of completely different origin and nature which cannot be separated experimentally. The diffraction here can be regarded as a kind  of "background" of the measurement.   

This fact might be very important when we are developing and using  other than partial wave  nuclear interaction model describing the nuclear elastic scattering.  Before we start to compare its prediction with experimental data we must be sure that this model includes also diffraction.  In nuclear elastic scattering channel (and only there)  the "interaction region" is not limited to the region where the nuclear forces act but due to the existence of quantum diffraction grating outside the nucleus extends far outside of it.

Quantum diffraction gratings existing outside the quantum objects itself might be an interesting additional tool for studying the properties of the waves of matter.

\end{document}